\documentclass[a4paper,preprintnumbers,floatfix,superscriptaddress,pra,twocolumn,showpacs,notitlepage]{revtex4-1}
\usepackage[utf8]{inputenc}
\usepackage[T1]{fontenc}
\usepackage[sc,osf]{mathpazo}
\usepackage{amsmath}
\usepackage{amsmath}
\usepackage{amssymb}
\usepackage{bbold}
\usepackage{graphicx}
\usepackage{ulem}
\usepackage{xcolor}
\usepackage{pgf,tikz}
\usepackage{enumerate}
\usepackage[colorlinks=true,citecolor=blue,urlcolor=blue]{hyperref}
\usepackage{diagbox}

\begin{document}

\title{Nonlocality Distillation and Quantum Voids}
\author{S. G. A. Brito}
\affiliation{International Institute of Physics, Federal University of Rio Grande do Norte, 59070-405 Natal, Brazil}
\author{M. G. M. Moreno}
\affiliation{International Institute of Physics, Federal University of Rio Grande do Norte, 59070-405 Natal, Brazil}
\author{A. Rai}
\affiliation{International Institute of Physics, Federal University of Rio Grande do Norte, 59070-405 Natal, Brazil}
\author{R. Chaves}
\affiliation{International Institute of Physics, Federal University of Rio Grande do Norte, 59070-405 Natal, Brazil}
\affiliation{School of Science and Technology, Federal University of Rio Grande do Norte, 59078-970 Natal, Brazil}

\date{\today}
\begin{abstract}
Via nonlocality distillation, a number of copies of a given nonlocal correlation can be turned into a new correlation displaying a higher degree of nonlocality. Apart from its clear relevance in situations where nonlocality is a resource, distillation protocols also play an important role in the understanding of information-theoretical principles for quantum theory. Here, we derive a necessary condition for nonlocality distillation from two copies and apply it, among other results, to show that $1$D and $2$D quantum voids --faces of the nonlocal simplex set with no quantum realization-- can be distilled up to PR-boxes. With that, we generalize previous results in the literature. For instance, showing a broad class of post-quantum correlations that make communication complexity trivial and violate the information causality principle. 
\end{abstract}
\maketitle

\section{Introduction}
Quantum correlations lie at the core of quantum enhanced information processing. Most prominently, entanglement \cite{RevModPhys.81.865} is a key ingredient in a variety of relevant applications, ranging from quantum computation \cite{nielsen2002quantum} to quantum metrology \cite{giovannetti2011advances} and quantum communication \cite{gisin2007quantum}. In the more recent years, another cornerstone in the foundations of quantum mechanics, Bell nonlocal \cite{RevModPhys.86.419} correlations have also been brought to the spotlight. As proved by John Bell in 1964 \cite{PhysicsPhysiqueFizika.1.195}, local measurements on distant entangled particles can generate correlations incompatible with any local realistic model, a result confirmed experimentally over and over \cite{hensen2015loophole,PhysRevLett.115.250402,PhysRevLett.115.250401} and of fundamental implication to our understanding of quantum theory. However, only more recently nonlocality has started to be seen as a resource. In a seminal paper \cite{PhysRevLett.67.661}, Ekert showed that the violation of a Bell inequality can be employed in quantum cryptography. This result has been brought to its extreme with the emergence of the device-independent framework to quantum information where the success of protocols is achieved without the need of a precise description of the underlying physical apparatuses. Within this context, nonlocality is now seen as resource in a number of applications beyond cryptography such as entanglement \cite{PhysRevLett.111.030501} and randomness \cite{acin2016certified} certification, self-testing \cite{Mayers:2004:STQ:2011827.2011830} and communication complexity problems \cite{PhysRevLett.92.127901}.

\begin{figure}[t!]
    \centering
    \includegraphics[scale=0.25, angle=270]{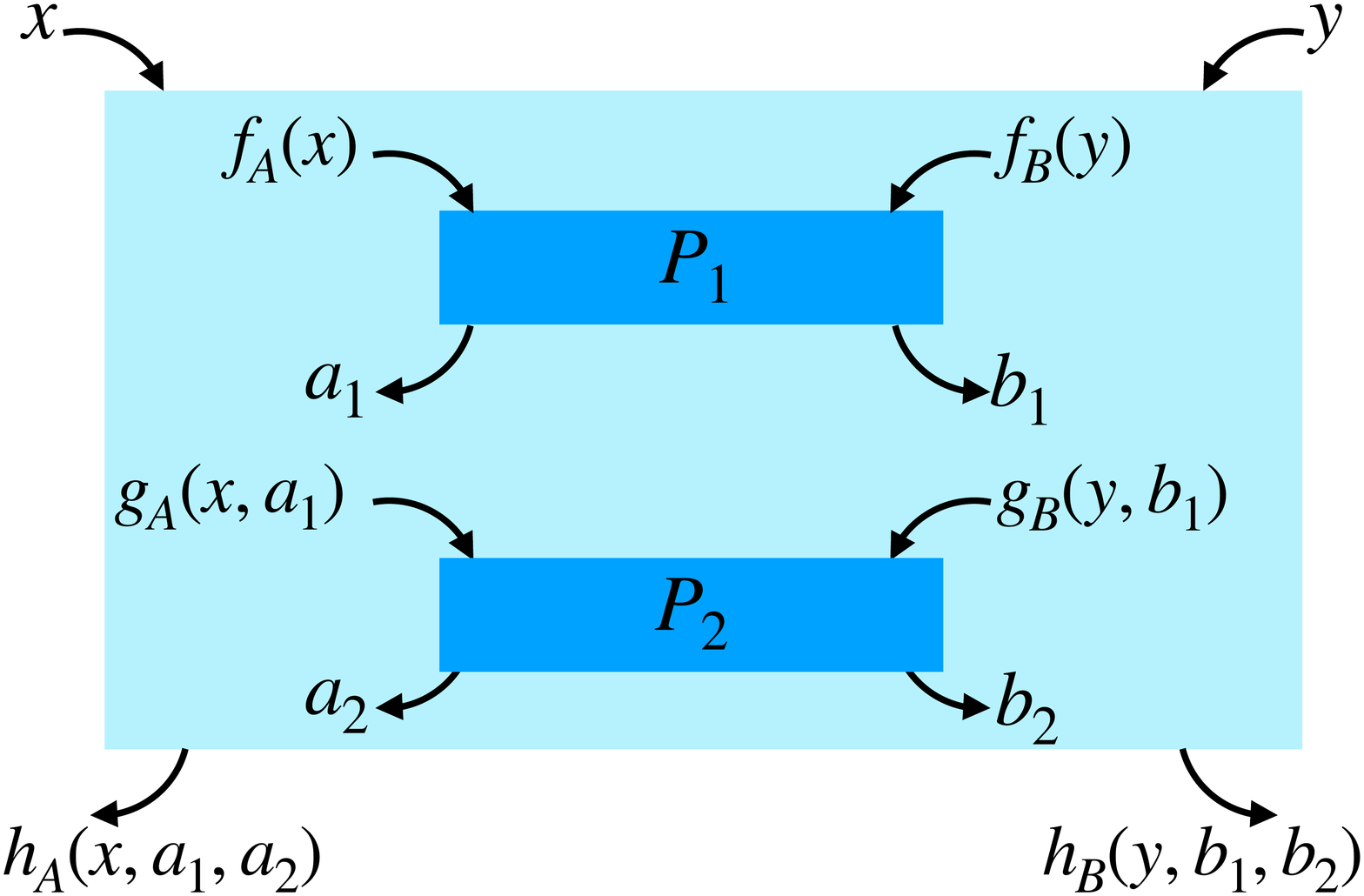}
    \caption{Illustration of a wiring protocol between two probability distributions (represented as boxes).}
    \label{wiring}
\end{figure}

As it happens to entanglement, nonlocal correlations typically become more useful the stronger they get. For instance, the more the paradigmatic CHSH inequality \cite{PhysRevLett.23.880} is violated the better is the bound we can put in the amount of randomness of a given measurement outcome \cite{acin2016certified}, the maximum being achieved exactly by maximum entangled states. In practice, however, due to noise and other uncontrollable source of errors, we often might have weak or not maximum entanglement or nonlocal correlations \cite{PhysRevA.89.042106,PhysRevA.86.012108}. To circumvent that, one has to rely on a distillation protocol: starting from two or more copies of a given resource, one can through a set of free operations extract a smaller number of copies but with more of the resource of interest. In the case of entanglement, these free operations are local operations and classical communication (since they cannot create or increase entanglement) \cite{RevModPhys.81.865}. In turn, the resource theory of nonlocality \cite{de_Vicente_2014} implies that such free operations are the so-called wirings \cite{PhysRevA.73.012101} (see Figure \ref{wiring}).

The first nonlocality distillation protocol has been introduced in \cite{forster} and, since then, a number of results have been obtained \cite{brunner,hoyer,PhysRevLett.106.020402,PhysRevA.80.062107,Lang_2014,PhysRevA.83.062114}, for instance showing the existence of bound nonlocality \cite{PhysRevLett.106.020402} and the fact that post-quantum correlations with a negligible amount of nonlocality can make communication complexity trivial \cite{brunner}. In spite of that, it is fair to say that few general conclusions have been obtained, typically referring to very specific classes of correlations. The difficulty relies on the fact that the number of possible wirings involved in a nonlocality distillation protocol increases very fast. As proven in \cite{PhysRevA.73.012101}, the set of possible protocols define a convex set, the extremal points of which are finitely many deterministic wirings. However, already at simplest possible Bell scenario, the CHSH scenario with two parties and two dichotomic measurements per party \cite{PhysRevA.89.042106}, there are  $82^4 = 45.212.176$ deterministic wirings, reason why we have seen slow progress in this research direction. 

On the more fundamental side, nonlocality distillation also plays a key role in the search for information-theoretical principles able to characterize the set of quantum correlations. It is known that special relativity alone cannot single out the quantum set, as there are correlation compatible with the non-signalling principle but beyond what can be achieved with quantum theory \cite{popescu1994quantum}. This has motivated the introduction of several new principles \cite{pawlowski2009information,chaves2015information,navascues2009glance,fritz2013local,navascues2015almost,van2013implausible,PhysRevLett.96.250401}. However, as noticed in \cite{PhysRevA.80.062107}, whatever principle a physical theory fulfills, the set of correlations it generates should be closed under wirings, implying non trivial constraints on the search for physical principles and the axiomatization of quantum theory \cite{Lang_2014}.  

In this paper we aim to provide somewhat more general statements on nonlocality distillation and their implications for information theoretical principles. Because of the difficulty mentioned above, we focus here on the CHSH scenario and distillation protocols involving two copies of the nonlocal correlations. Within this context, we first obtain a general necessary condition for nonlocality distillation with two copies. Then we employ it to analyze faces of the nonlocal simplex of correlations with no quantum realization, the so called quantum voids \cite{PhysRevA.99.032106} (if the dimension of the face is $k$ then the quantum void is said to be $k$-dimensional. Using our necessary condition we prove that correlations in all $1$D and some of the $2$D quantum voids are distillable to maximal nonlocality. This allows us to generalize previous results in the literature. First, we show that there are whole faces of the non-signalling set violating the principle of non-trivial communication complexity. Finally, we show how a large class of correlations not violating Uffink's inequality (a necessary condition for a correlation to be compatible with the principle of information causality) can do so after a distillation protocol.

The paper is organized as follows. In Section \ref{sec:tool} we introduce the basic concepts and tools necessary to state our results. In Section \ref{sec:NLdist} we propose a necessary condition for nonlocality distillation, that is then employed in Section \ref{sec:NLqvoid} to prove that all $1$D quantum voids and some $2$D and $3$D quantum voids correlations are distillable. In Sections \ref{sec:CC} and \ref{sec:IC}  we look for the consequences of distillation protocols in the non-trivial communication complexity \cite{van2013implausible} and information causality principles \cite{pawlowski2009information}. Finally, in Section \ref{sec:discussion} we present our conclusions. 

\section{Toolbox}
\label{sec:tool}
We will restrict our attention here to the simplest possible Bell scenario, also known as the Clauser-Horne-Shimony-Holt (CHSH) scenario \cite{PhysRevLett.23.880}. It refers to two spatially separated parties, Alice and Bob, which upon receiving their shares of a joint physical system randomly and independently choose one of two possible dichotomic measurements to perform. Their measurement choices are labelled by the random variables $X$ and $Y$ and the measurement outcomes as $A$ and $B$, for Alice and Bob respectively. In the CHSH scenario all these variables are binary, that is, $x,y,a,b=0,1$. The results of a Bell experiment are encoded in a probability distribution of these variables $p(a,b,x,y)$ and since typically we do have control over the input distribution $p(x,y)$ it has become customary to work with the conditional probability distribution $p(a,b \vert x,y)$. Within this context, the fundamental question is to understand what are the constraints implied by different kind of theories on the possible correlations that can be observed in a Bell experiment.

Under the assumption known as local realism, the probability distribution should be decomposable as 
\begin{equation}
\label{pc}
p(a,b \vert x,y) = \sum_{\lambda}p(\lambda) p(a \vert x, \lambda) p(b \vert y, \lambda),
\end{equation}
defining a set of correlation $\mathcal{L}$ where we have explicitly imposed the following conditions. First, the ``realism'' condition implying the existence of a hidden variable $\Lambda$ determining the probabilities of measurement outcomes even of measurement choices that have not been performed. Second, the locality condition stating that only variables in the causal past of the measurement outcomes might have a causal influence over their statistics, that is, $p(a\vert x,y,b,\lambda)= p(a\vert x,\lambda)$ (similarly for Bob). Finally, we also have to impose the measurement independence assumption (also known as ``free will'') implying that $p(x,y,\lambda)=p(x,y)p(\lambda)$.

In turn, the quantum description for this experiment, implies via the Born rule that the distribution should be written as
\begin{equation}
\label{pq}
p(a,b \vert x,y) = \mathrm{Tr}\left[\left(M_a^x \otimes M_b^y \right) \rho \right],
\end{equation}
defining a set of correlations $\mathcal{Q}$ where $M_a^x$ and $M_b^y $ describe measurement operators and $\rho$ is the density operator describing the joint physical system shared by Alice and Bob. Bell's theorem \cite{PhysicsPhysiqueFizika.1.195} shows that there are quantum distributions of the form \eqref{pq} incompatible with the classical description given by \eqref{pc}, the phenomenon known as quantum nonlocality that can be witnessed by the violation of the CHSH inequality \cite{PhysRevLett.23.880}
\begin{equation}
\label{chsh}
\mathrm{CHSH}= E_{00}+E_{01}+E_{10}-E_{11} \leq 2    
\end{equation}
that in quantum case can achieve $\mathrm{CHSH}= 2 \sqrt{2}$ where $E_{xy}=P(a=b \vert xy)-P(a \neq b \vert xy)$.

A third possible description is to wonder what are the implications on the correlation arising from imposing special relativity to this Bell experiment. Because of the space-like separation between the parties, we see that the statistics observed locally by one of the parties should be completely independent of whatever choice of measurement the other party is doing, otherwise they could communicate superluminally. Mathematically, this is described by following set of linear constraints on the probabilities, known as non-signalling (NS) conditions:
\begin{eqnarray}
\label{pns}
& & p(a\vert x) = \sum_{b} p(a,b\vert x,y) =  \sum_{b} p(a,b\vert x,y^{\prime}) \\ \nonumber
& & p(b\vert y) = \sum_{a} p(a,b\vert x,y) = \sum_{a} p(a,b\vert x^{\prime},y),
\end{eqnarray}
defining a set of correlations $\mathcal{NS}$. Strikingly, there are non-signalling correlation beyond what can be achieved with quantum theory \cite{popescu1994quantum}. In short, we know that these $3$ sets of correlations respect the following strict inclusion relation: $\mathcal{L} \subsetneq \mathcal{Q} \subsetneq \mathcal{NS}$.

The sets $\mathcal{L}$ and $\mathcal{NS}$ are polytopes, convex sets described by a finite number of extremal points or equivalently a finitely many facets (linear inequalities). In the CHSH scenario, the extremal points of the $\mathcal{NS}$ set have been fully characterized \cite{PhysRevA.71.022101}, consisting of 8 extreme nonlocal points and 16 extreme local points described below. In a Bell scenario we restrict our attention to the probability distribution, regardless of the internal working of the measurement and state preparation devices. For this reason, it is typical to refer to correlations (alternatively, probability distributions) simply as boxes.

\begin{itemize}
  \item \textbf{PR-box $PR(ab|xy)$}: a no-signaling correlation that maximally violates the CHSH inequality \eqref{chsh} or one of its symmetries. There are 8 of such boxes:
    \begin{equation}
        \nonumber
        PR^{\mu\nu\sigma}(ab|xy)=\frac{1}{2}\delta_{a\oplus b,xy\oplus\mu x\oplus\nu y\oplus \sigma},
    \end{equation}
    where $\delta_{[.],[.]}$ representing the Dirac's delta.
    
    \item \textbf{Local-box $L_{\alpha\beta\gamma\theta}$}: there are 16 deterministic local boxes that can be parametrized as:
    \begin{equation}
    \nonumber
    L_{\alpha\beta\gamma\theta}(ab|xy)=\delta_{a,\alpha x\oplus\beta}\delta_{b,\gamma y\oplus \theta}.
    \end{equation}
\end{itemize}

In the CHSH scenario, any nonlocal distribution can be decomposed as the convex sum of a single PR-box plus up to eight more local deterministic strategies. For instance, any correlation violating the CHSH inequality \eqref{chsh} can be written as 
\begin{equation}
    \label{defbox}
        p(ab|xy)=c_0 PR(ab|xy) + \sum_{\alpha,\beta,\gamma=0}^{1} c_{\alpha\beta\gamma} L_{\alpha\beta\gamma}(ab|xy),
    \end{equation}
    where
    \begin{equation}
    \label{defPR}
    PR(ab|xy)=PR^{000}(ab,xy),
    \end{equation}
    \begin{equation}
    \label{deflocal}
    L_{\alpha\beta\gamma}(ab|xy)=L_{\alpha\beta\gamma(\alpha \gamma\oplus \beta)}(ab|xy),    
    \end{equation}
and $c_0 + \sum_{\alpha,\beta,\gamma=0}^{1} c_{\alpha\beta\gamma}=1$, $0\leq c_0\leq 1$ and $0\leq c_{\alpha\beta\gamma}\leq 1$ $\forall$ $\alpha$, $\beta$, and $\gamma$. In this case $\mathrm{CHSH}(PR)=4$ (maximal violation) and $\mathrm{CHSH}(L_{\alpha\beta\gamma})=2$ (these local points saturate the local bound of the inequality). Furthermore, any other nonlocal distribution can be achieved via local reversible transformations (relabelings) over such distribution. In this sense, in the CHSH scenario it is thus sufficient to consider only \eqref{defbox} and we will do so in what follows. To simplify the notation, from now on we will refer to the local deterministic strategies as $L_i$ (where might assume eight different values $i=1,\dots,8$). Here, we follow the notation in \cite{PhysRevA.99.032106}, as shown in Table \ref{Li_Labg}.

\begin{table}[h!]
    \centering
    $\begin{array}{|c|c|c|c|c|c|c|c|c|}
 \hline
 L_i & L_1 & L_2  & L_3 &L_4 & L_5 & L_6 & L_7& L_8  \\
\hline
 L_{\alpha\beta\gamma} & L_{101} & L_{111} & L_{001} & L_{011} & L_{110} & L_{100} & L_{000} & L_{010} \\
 \hline
\end{array}$
    \caption{The correspondence between $L_i$ and  $L_{\alpha\beta\gamma}$.}
    \label{Li_Labg}
\end{table}

As will be described in more details below, we are interested here in nonlocality distillation. That is, starting with two copies with a certain degree of nonlocality we want that the final wired correlation has a higher nonlocality degree. For that, we first have to define a quantifier. Different measures have been considered before, the violation of the CHSH inequality itself \cite{forster} and the so called EPR-$2$ decomposition \cite{PhysRevLett.106.020402}. Here we employ the trace distance measure introduced in \cite{PhysRevA.97.022111}, basically quantifying the minimum distance of the nonlocal point in question to the set of local correlations. In the CHSH scenario the trace distance quantifier has been shown to be equivalent to the CHSH inequality violation (up to a constant factor), reason why we consider here as a quantifier $NL(p)$ of the nonlocality of a given distribution $p=p(ab\vert xy)$ the following quantity:
    \begin{equation}
    NL(p)=
            \max\left[ \frac{\Pi(\mathrm{CHSH})-2}{2}, 0 \right] 
    \end{equation}
where $\Pi(CHSH)$ stand for all eight simmetries of the CHSH inequality, thus bounding $NL=0$ for local points and $NL=1$ for the PR-box and its symmetries (maximal nonlocality).

\subsection*{Faces of the nonlocal set and quantum voids} \label{nlsimplex}

 Following \cite{PhysRevA.99.032106}, we represent a probability distribution $p(ab\vert xy)$ as a vector $(p_1,p_2,...,p_{16})$, where the ordering of probabilities is as shown in Table~\ref{tab2}.

 \begin{table}[h]
 \parbox{.35\linewidth}{
	\centering
	\begin{tabular}{|c||c|c|c|c|}
		\hline
		\diagbox[width=3em]{$\mathbf{xy}~$}{$\mathbf{~ab}$}& $00$&$01$&$10$& $11$\\ \hline \hline
		
		$00$& $p_1$&\color{blue}{$p_{2}$}&\color{blue}{$p_{3}$}& $p_{4}$\\ \hline
		$01$& $p_{5}$&\color{blue}{$p_{6}$}&\color{blue}{$p_{7}$}& $p_{8}$\\ \hline
		$10$& $p_{9}$&\color{blue}{$p_{10}$}&\color{blue}{$p_{11}$}& $p_{12}$\\ \hline
		$11$& \color{blue}{$p_{13}$}&$p_{14}$&$p_{15}$& \color{blue}{$p_{16}$}\\ \hline
		
	\end{tabular} \label{tab2}}
	\hfill
	\parbox{.45\linewidth}{
	\centering
	\begin{tabular}{|c||c|c|c|c|}
		\hline
		\diagbox[width=3em]{$\mathbf{xy}~$}{$\mathbf{~ab}$}& $00$&$01$&$10$& $11$\\ \hline \hline
		
		$00$& $1/2$&\color{blue}{$0$}&\color{blue}{$0$}& $1/2$\\ \hline
		$01$& $1/2$&\color{blue}{$0$}&\color{blue}{$0$}& $1/2$\\ \hline
		$10$& $1/2$&\color{blue}{$0$}&\color{blue}{$0$}& $1/2$\\ \hline
		$11$& \color{blue}{$0$}&$1/2$&$1/2$& \color{blue}{$0$}\\ \hline
		
	\end{tabular}}
    
\caption{Joint probabilities $p(ab\vert xy)$ are ordered as shown in the first table. The free variables corresponds to those that are zero in $PR^{000}$ (second table).}
\end{table}

 On considering the probabilities $p_2$, $p_3$, $p_6$, $p_7$, $ p_{10}$, $p_{11}$, $p_{13}$ and $p_{16}$ as free variables, the remaining eight probabilities can be expressed in terms of the free variables by using no-signaling and normalization conditions. Notice that the free variables corresponds to probabilities taking value zero in PR-box $(PR^{000})$ correlation, and for the local box $L_i$ the free variable probabilities are such that $p_k = 1$ for the correspondent free variable and zero for the other seven, where $k$ is the index of the correspondent free variable. In this way, we related  $L_1\rightarrow p_2$, $L_2\rightarrow p_3$, $L_3\rightarrow p_6$, $L_4\rightarrow p_7$, $L_5\rightarrow p_{10}$, $L_6\rightarrow p_{11}$, $L_7\rightarrow p_{13}$ and $L_8\rightarrow p_{16}$. 

 We are interested in the region which is convex hull of the PR-box and eight local vertices $\{L_i: 1\leq i\leq 8\}$, which forms an eight dimensional simplex that we refer as the nonlocal simplex ($NLS$). In particular, we will consider the faces of $NLS$, where given a convex set $C\subseteq R^n$ and a supporting hyperplane $H$ of $C$, the set of points in $H \cap C$ defines a face of $C$ \cite{boyd2004convex}. Further, we will consider nonlocal faces of the region $NLS$, and all such faces can be derived by setting some of the free variable probabilities to zero.
 
As shown in \cite{PhysRevA.99.032106}, nonlocal faces can give rise to quantum voids, faces where all nonlocal points are of a postquantum nature. Non-signalling faces of dimension four and smaller are all quantum voids, as well as some of the faces of dimension five and six. These are the sets we will focus throughout out the paper.
 
\section{Nonlocality distillation and wirings} 
\label{sec:NLdist}
Representing the correlations as a box with inputs and outputs, a nonlocality distillation protocol can be basically understood as a wiring among two boxes, where for instance the outcomes of the first box can be used as the input for the second one. If we wire two distant distributions as $W(p(ab|xy),p^{\prime}(ab|xy))$ we obtain a new distribution $q(ab|xy)$ given by:
         \begin{eqnarray}
         \label{defwiring}
              & & q(ab|xy) = W(p(ab|xy),p^{\prime}(ab|xy))\\
             \nonumber
             & & =\sum_{a_1,a_2,x_1,x_2,b_1,b_2,y_1,y_2=0}^{1} \chi_x(a,a_1,a_2,x_1,x_2)\chi_y(b,b_1,b_2,y_1,y_2)\cdot \\ \nonumber
             & &\cdot p(a_1b_1|x_1y_1)p^{\prime}(a_2b_2|x_2y_2)
         \end{eqnarray}
where $\chi_x(a,a_1,a_2,x_1,x_2)$ represents the wiring performed locally by Alice and $\chi_y(b,b_1,b_2,y_1,y_2)$ the wiring performed by Bob. Here $x_i$ and $y_j$ are respectively $i$th and $j$th inputs to Alice and Bob's boxes, and $a_i$ and $b_j$ are the outcomes from respective boxes, $a$ and $b$ are the final respective outputs of Alice and Bob; since two boxes are wired, $i,j \in\{1,2\}$. All possible wirings for boxes with possible inputs and outputs have been characterized \cite{PhysRevA.73.012101} and  form a convex set whose vertices are described according to five different classes in the Table below.
 \begin{widetext}  
    \begin{center}
        \begin{tabular}{| c | c |}
        \hline
         Potential couplers classes & $\chi(a,a_1,a_2,x_1,x_2)=1$ ($0$ otherwise) \\
         \hline
         $\chi_{\mu}^D$ & $x_1=x_2$ and $a=\mu$         \\
         \hline
         $\chi_{\mu\nu\sigma}^O$ & $x_1=x_2=\mu$ and $a=a_{\nu+1}\oplus\sigma$\\
         \hline
         $\chi_{\mu\nu\sigma}^X$ & $x_1=\mu$, $x_2=\nu$, and $a=a_1\oplus a_2\oplus\gamma$\\
         \hline
         $\chi_{\mu\nu\sigma\delta\epsilon}^A$ & $x_1=\mu$, $x_2=\nu$, and $a=(a_1\oplus\sigma)(a_2\oplus\delta)\oplus\epsilon$
         \\
         \hline
         $\chi_{\mu\nu\sigma\delta\epsilon}^S$ & $x_{\mu+1}=\nu$, $x_{(\mu\oplus 1)+1}=a_{\mu+1}\oplus\sigma$, and $a=a_{(\mu\oplus 1)+1}\oplus\delta a_{\mu+1}\oplus\epsilon$\\
         \hline
    \end{tabular}
    \end{center}
\end{widetext} 

Before moving on to derive a necessary criterion for nonlocality distillation, let us mention that in the CHSH scenario \cite{PhysRevLett.23.880}, some examples of sets closed under wirings are known. The set of local correlations, quantum correlations, and no-signaling correlations as well as several sets of correlations between local and quantum sets as well as  all sets of correlations generated by different levels of the NPA hierarchy~\cite{NPA}. See Ref.~\cite{Lang_2014} for a discussion about sets of correlations closed under wirings. Now since some of these sets are very closely spaced (for example, NPA hierarchies), for points in these regions it is very hard to come up with distillation protocols; the only viable option is to find the wirings which generates a flow gazing on the boundary of the sets closed under wirings, which turns out to be a difficult task \cite{Lang_2014}.

Considering two copies of an initial box $p(ab\vert xy)$ given by \eqref{defbox}, after the wiring we obtain
\begin{align}
\nonumber
    W(p,p)&=c_0^2 B_0^0+c_0\sum_{i=1}^{8}c_{i}(B^0_{i}+B_0^{i})\\
    &+\sum_{i,j=1}^{8}c_{i}c_{j}B^{i}_{j},
\end{align}
where
\begin{eqnarray}
\label{Bs}
 B_0^0(ab|xy)= & & W(PR(ab|xy),PR(ab|xy)) \\ \nonumber
    B_{i}^0(ab|xy)= & & W(PR(ab|xy),L_{i}(ab|xy)) \\ \nonumber
    B_0^{i}(ab|xy)= & &W(L_{i}(ab|xy),PR(ab|xy)) \\ \nonumber
    B^{i}_{j}(ab|xy)= & &W(L_{i}(ab|xy),L_{j}(ab|xy))
\end{eqnarray}

The nonlocality of the initial distribution is $NL(p)=c_0$, since $NL$ is linear for the distribution \eqref{defbox} and $NL(PR)=1$ and $NL(L_{i})=0$ for all $i$. For the wired distribution we obtain directly an upper bound for its nonlocality, given by
\begin{eqnarray}
\label{NLwired}
    & &NL(W(p,p)) \leq c_0^2 NL(B^0_0) \\ \nonumber & & +c_0\sum_{i=1}^{8}c_{i}(NL(B^0_{i})+NL(B_0^{i})).
\end{eqnarray}
Clearly, to have nonlocality distillation we need $NL(W(p,p)) > NL(p)$ thus implying that
\begin{eqnarray}
& & c_0NL(B^0_0) \\ \nonumber 
& & +\sum_{i=1}^{8}c_{i}(NL(B^0_{i})+NL(B_0^{i})) > 1. 
\end{eqnarray}

Since $c_0+\sum_{i=1}^{8}c_{i}=1$, the condition for distillation becomes
\begin{equation}
    c_0(NL(B_0 ^0)-1)+\sum_{i=1}^{8}c_{i}(NL(B^0_{i})+NL(B^{i}_0)-1)>0
\end{equation}
As hinted by the expression above, looking at the nonlocality of the wired terms in \eqref{Bs} can provide us with the necessary information to decide whether a given correlation is distilablle or not, something we will explore in what follows. 

By testing all $82^4$ possible deterministic wirings on the wired distributions $B^0_0$, $B^{i}_0$ and $B_{i}^0$ one can prove that $NL(B^0_0)=\left\{0,1/2,1 \right\}$ and $NL(B^{i}_0)=\left\{0,1 \right\}$ as well as $NL(B_{i}^0)=\left\{0,1 \right\}$. The necessary condition above then reduces to
\begin{equation}
\label{necessary}
    c_0(NL(B_0^0)-1)+\sum_{i=0}^{2}\tilde{c}_{i}(i-1)>0
\end{equation}
where $\sum_{i=0}^{2}\tilde{c}_{i}=\sum_{k=1}^{8} c_{k}$ and $\tilde{c}_i$ is the sum of all coefficients $c_{k}$ for which $NL(B^0_{k})+NL(B^{k}_0)=i$.
Notice that this necessary condition can be raised to a sufficient one if one can guarantee that every term in the initial distribution \eqref{defbox} is being mapped under the wiring to a distribution of the same form \eqref{defbox}.

\section{Nonlocality distillation in quantum voids}
\label{sec:NLqvoid}
In what follows we are going to consider whether entire faces of the nonlocal simplex (in particular, quantum voids) are distillable or not. In this case, since the coefficient $c_0$ in \eqref{defbox} can vary as $0 < c_0 < 1$, a necessary condition for distillation is that $NL(B_0 ^0)=1$ (remember that only assume 3 possible values $NL(B_0 ^0)=\left\{0,1/2 ,1 \right\}$). Furthermore, without loss of generality, we can restrict to those wirings such that $B_0 ^0=PR(ab \vert xy)$. Making this restriction, reduces the number of wirings from $82^4$ to $3152$, making a complete analysis of the wirings amenable. Furthermore, in this case, the necessary condition for distillation is then simply given by
\begin{equation}
\label{face_condition}
    \tilde{c}_2>\tilde{c}_0.
\end{equation}

\subsection{All correlations in a $1$D quantum void can be distilled to a PR-box}
Consider a generic $1$D quantum void described by the distribution
\begin{equation}
p=c_0 PR+ (1-c_0)L_{i}.  
\end{equation}
In this case, the condition \eqref{face_condition} simply states that there should exist at least one wiring for which $\tilde{c}_2=1$. Furthermore, this becomes a sufficient condition as well if the terms $B_{i}^{0}$ and $B_{0}^{i}$ are indeed mapped to $PR$. By searching over the $3152$ wirings, we found that for every local point $L_{i}$ there is a strategy doing that (see Table \ref{1dvoid_strategies}).
That is, every point in a $1$D quantum void is distillable and the nonlocality of the wired correlation is raised from $NL(p)=c_0$ to $NL(W(p,p))=c_0(2-c_0)$. 

\begin{table}[h!]
    \centering
$\begin{array}{ |c|c|c|c|c|} 
 \hline
 L_i & x=0 & x=1 & y=0 & y=1 \\ 
 \hline
L_1 & \chi ^S{}_{0,0,0,0,0} & \chi ^X{}_{1,1,1} & \chi ^S{}_{0,0,1,0,0} & \chi ^X{}_{1,1,0} \\
 L_2 & \chi ^S{}_{0,0,1,0,0} & \chi ^X{}_{1,1,0} & \chi ^S{}_{0,0,0,0,0} & \chi ^X{}_{1,1,1} \\
 L_3 & \chi ^S{}_{0,0,0,1,0} & \chi ^X{}_{1,1,0} & \chi ^X{}_{0,0,0} & \chi ^S{}_{0,1,0,0,0} \\
 L_4 & \chi ^S{}_{0,0,1,1,1} & \chi ^X{}_{1,1,1} & \chi ^X{}_{0,0,1} & \chi ^S{}_{0,1,1,0,0} \\
 L_5 & \chi ^X{}_{0,0,1} & \chi ^S{}_{0,1,1,0,0} & \chi ^S{}_{0,0,1,1,1} & \chi ^X{}_{1,1,1} \\
 L_6 & \chi ^X{}_{0,0,0} & \chi ^S{}_{0,1,0,0,0} & \chi ^S{}_{0,0,0,1,0} & \chi ^X{}_{1,1,0} \\
 L_7 & \chi ^X{}_{0,0,0} & \chi ^S{}_{0,1,1,1,0} & \chi ^X{}_{0,0,0} & \chi ^S{}_{0,1,1,1,0} \\
 L_8 & \chi ^X{}_{0,0,1} & \chi ^S{}_{0,1,0,1,1} & \chi ^X{}_{0,0,1} & \chi ^S{}_{0,1,0,1,1} \\
 \hline
\end{array}$
\caption{$1$D quantum void. Above we show one example of a strategy that can take any point in the set $(PR,L_{i})$ up to PR. There exist at least $8$ different strategies for each set that can do this. Using the same strategy more or less $10$ times we can increase the non locality of any point maximally (to a PR-box).}
\label{1dvoid_strategies}
\end{table}

Furthermore, as we will show below, not only every $1$D quantum void is distillable, but can as well be asymtotically distilled (by the iterative application of the same wiring) to a PR-box. Notice, that this does not follow directly from the fact that every nonlocal correlation in a $1$D quantum void is distillable. For instance, there are wirings that distill the nonlocality of the correlations but map them out of the $1$D void. However, if we can find a wiring that distill nonlocality keeping the correlation in the $1$D quantum void then we can guarantee violation up to a PR-box.

For example, for the one-dimensional face associated with $L_7$, using the wiring described in table \ref{1dvoid_strategies} we have that $B_0^0=PR$, $B_0^7=PR$, $B_7^0=PR$ and $B_7^7=L_7$. Thus, 
\begin{eqnarray}
W(p,p)=(1-(1-c_0)^2)PR + (1-c_0)^2L_7
\end{eqnarray}
Now, if we wire two boxes of the form $p_m=(1-(1-c_0)^{2^{m-1}})PR + (1-c_0)^{2^{m-1}} L_7$ we get:

\begin{eqnarray}
q=(1-(1-c_0)^{2^m})PR+(1-c_0)^{2^m}L_7
\end{eqnarray}
So we can use the principle of induction, to assure that, starting with a box $p=c_0PR+(1-c_0)L_7$, after $n$ interactions of wirings, we have the box $Q_n$:
\begin{eqnarray}
q_n=(1-(1-c_0)^{2^n})PR + (1-c_0)^{2^n}L_7
\end{eqnarray}
From that, we have that $q_n\rightarrow PR$ as $n\rightarrow\infty.$

\subsection{All correlations in some $2$D quantum voids can be distilled to a PR-box}
\begin{figure}[t!]
\begin{center}
\includegraphics[scale=0.35]{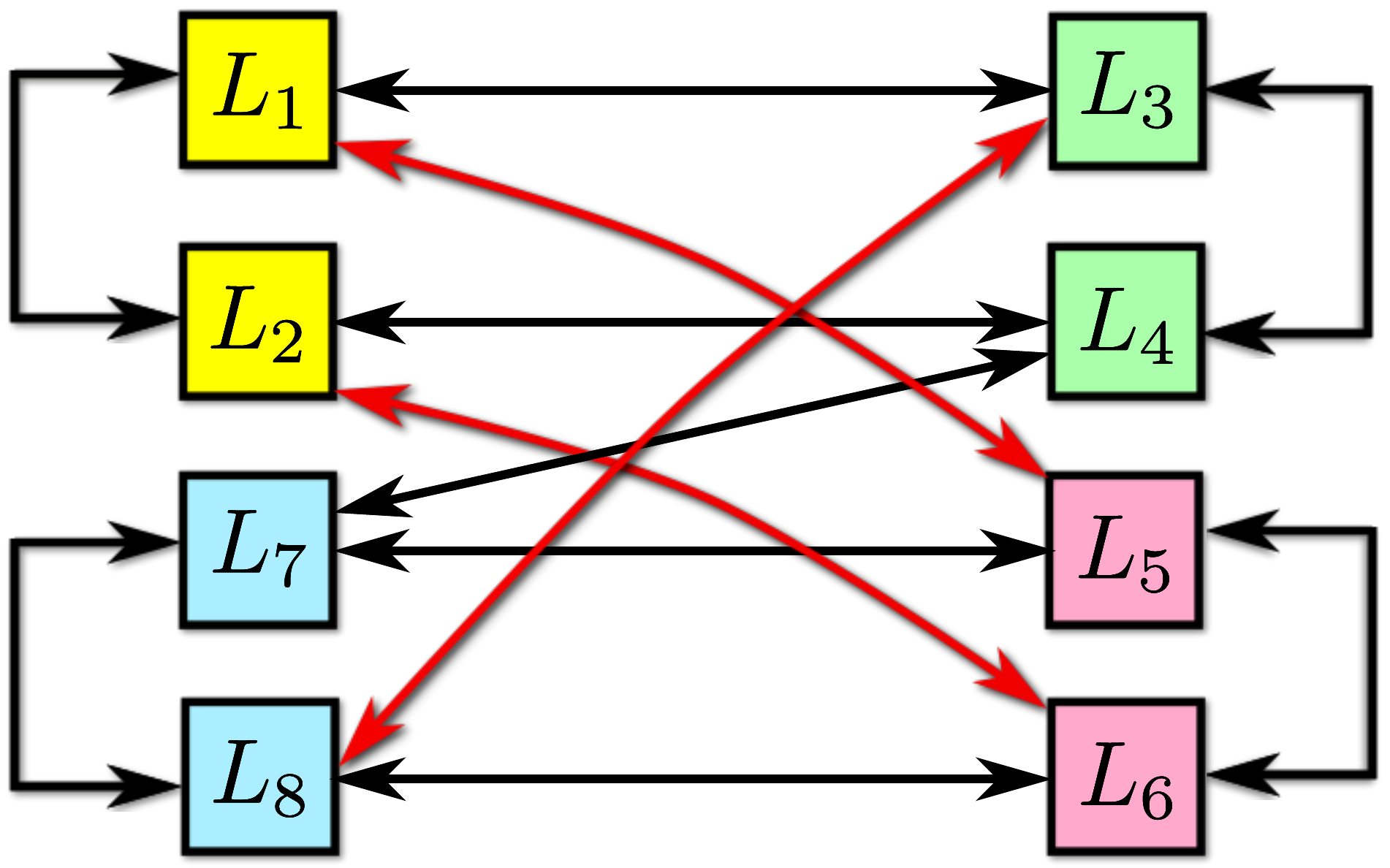}
\end{center}
\caption{Schematic representation of the $2$D quantum voids that are distillable up to a PR-box. If there exists one edge between two local vertices, then this set is completely distillable and there is at least one single strategy that can distill the entire void. For instance, the set $\{PR,L_1,L_2\}$ is completely distillable, however the set $\{PR,L_1, L_7\}$ is not.}
\label{2d_voids_distillable}
\end{figure}

Consider a generic $2$D quantum void described by the distribution
\begin{equation}
p=c_0 PR+ (1-c_0)\left( c_{i}L_{i} +c_{j}L_{j} \right) .  
\end{equation}
where we can order the coefficients as $c_i \geq c_j$. By searching over the $3152$ wirings such that $B^0_0=PR$ we found that there is always at most one value of $i$ (for all $i$) such that $B^0_{i}=B_0^{i}=PR$. Furthermore, there always exist a wiring such that $B^0_{j}$ and $B_0^j$ are mapped to a distribution of the form \eqref{defbox}, that is, the condition \eqref{defbox} becomes a sufficient condition for distillation. Thus, choosing a wiring such that $\tilde{c}_2=c_i$ we always will distill the nonlocality of the distribution, unless $\tilde{c_0}=c_j$ and $c_i=c_j$ (that is, we are at the isotropic line defined by the non-signalling facet). If $c_i=c_j$ we have to guarantee that $\tilde{c_0}=0$, what only happen at a subset of $2$D quantum voids given by the sets consisting of the following pairs of local points:
$(L_1L_2)$, $(L_1L_3)$,  $(L_1L_5)$, $(L_2L_4)$, $(L_2L_6)$, $(L_3L_4)$, $(L_3L_8)$, $(L_4L_7)$, $(L_5L_6)$, $(L_5L_7)$,  $(L_6L_8)$ and $(L_7L_8)$. All the other $2$D quantum voids are distillable but not at the isotropic line (see Fig. \ref{2d_voids_distillable}).

As a matter of fact, not only all these $2$D quantum voids are fully distillable but can as well be distilled up to a PR-box. In Table \ref{2dvoid_strategies} we show the example of a wiring strategy for each $2$D quantum void that being applied iteratively can bring any nonlocal correlation up to the PR-box.

\begin{table}[h!]
    \centering
$\begin{array}{ |c|c|c|c|c|} 
 \hline
  L_iL_{j} & x=0 & x=1 & y=0 & y=1 \\ 
 \hline
L_1L_2 & \chi ^S{}_{0,0,0,0,0} & \chi ^X{}_{1,1,1} & \chi ^S{}_{0,0,1,0,0} & \chi ^X{}_{1,1,0} \\
 L_1L_3 & \chi ^S{}_{0,0,0,0,0} & \chi ^X{}_{1,1,0} & \chi ^S{}_{0,1,1,0,0} & \chi ^S{}_{0,0,1,0,0} \\
L_1L_5 & \chi ^S{}_{0,0,0,0,0} & \chi ^S{}_{0,1,0,0,0} & \chi ^S{}_{0,0,1,0,0} & \chi ^X{}_{1,1,0} \\
L_2L_4& \chi ^S{}_{0,0,1,0,0} & \chi ^X{}_{1,1,0} & \chi ^S{}_{0,0,0,0,0} & \chi ^S{}_{0,1,0,0,0} \\
L_2L_6 & \chi ^S{}_{0,0,1,0,0} & \chi ^S{}_{0,1,1,0,0} & \chi ^S{}_{0,0,0,0,0} & \chi ^X{}_{1,1,1} \\
L_3L_4 & \chi ^S{}_{0,0,0,1,0} & \chi ^X{}_{1,1,0} & \chi ^X{}_{0,0,0} & \chi ^S{}_{0,1,0,0,0} \\
L_3L_8 & \chi ^S{}_{0,0,0,1,0} & \chi ^S{}_{0,1,1,1,0} & \chi ^X{}_{0,0,0} & \chi ^S{}_{0,1,0,0,0} \\
 L_4L_7& \chi ^S{}_{0,0,1,1,1} & \chi ^S{}_{0,1,0,1,1} & \chi ^X{}_{0,0,1} & \chi ^S{}_{0,1,1,0,0} \\
L_5L_6 & \chi ^X{}_{0,0,0} & \chi ^S{}_{0,1,1,0,1} & \chi ^S{}_{0,0,1,1,0} & \chi ^X{}_{1,1,0} \\
L_5L_7 & \chi ^X{}_{0,0,0} & \chi ^S{}_{0,1,1,0,0} & \chi ^S{}_{0,1,0,1,0} & \chi ^S{}_{0,0,1,1,0} \\
L_6L_8 & \chi ^X{}_{0,0,0} & \chi ^S{}_{0,1,0,0,0} & \chi ^S{}_{0,0,0,1,0} & \chi ^S{}_{0,1,1,1,0} \\
L_7L_8 & \chi ^X{}_{0,0,0} & \chi ^S{}_{0,1,1,1,0} & \chi ^X{}_{0,0,0} & \chi ^S{}_{0,1,1,1,0} \\ 
\hline
\end{array}$
\caption{$2$D quantum void. Above we show one example of a strategy that can take any point in the set $\{PR,L_i,L_j\}$ up to PR. From $28$ $2$D quantum voids only $12$ has this property. Using that strategies more or less $20$ times we can increase the non locality of any point maximally.}
\label{2dvoid_strategies}
\end{table}

\begin{figure}[t!]
\begin{center}
\includegraphics[scale=.6]{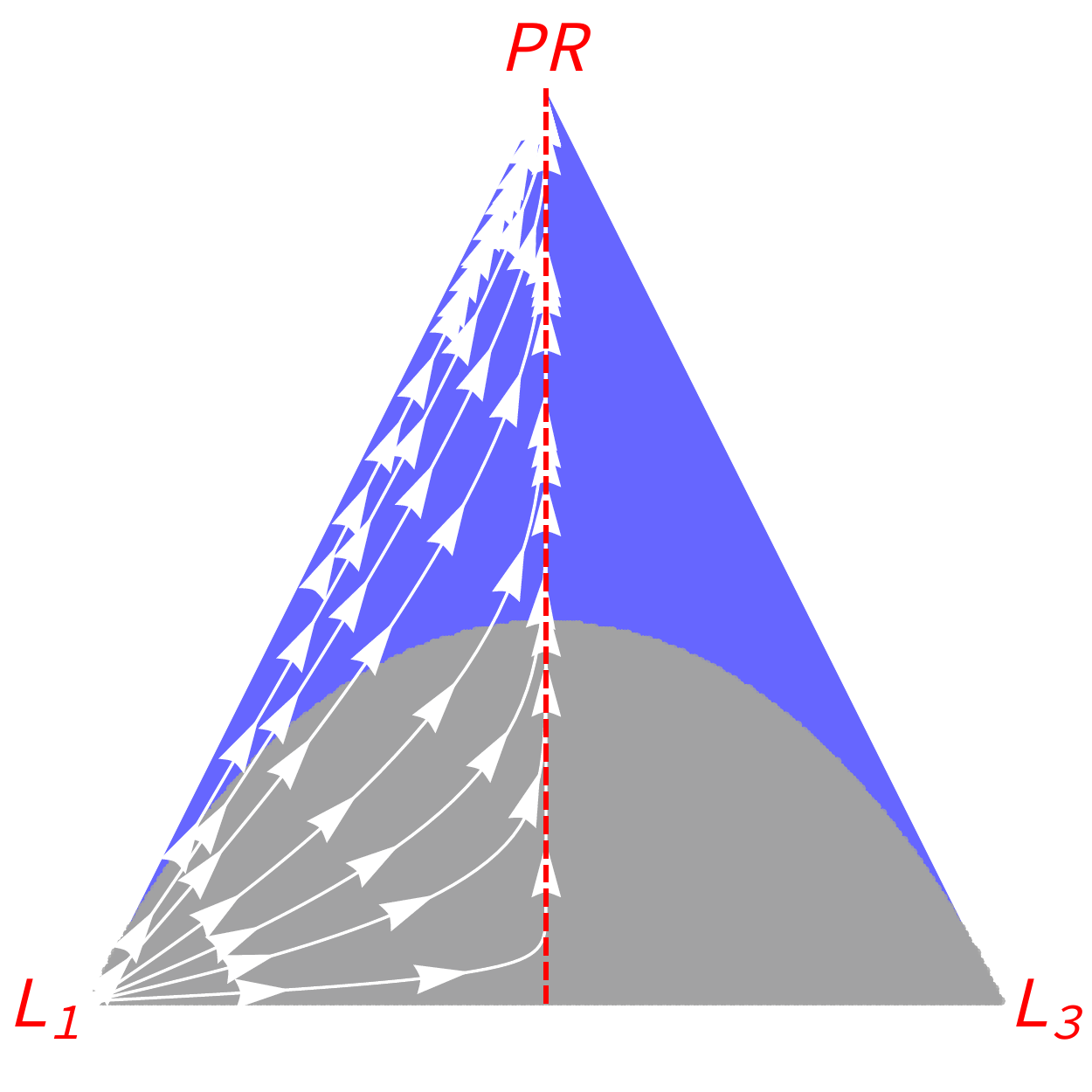}
\end{center}
\caption{Flow map for the distillation in the $2$D quantum void $\{PR,L_1,L_3\}$. By applying successively the wiring strategy shown in Table \ref{1dvoid_strategies} any point in this set is distilled to PR-box. Interestingly, the flow map shows that the wiring always moves the points up to the isotropic line (dashed red line) and then go up to the PR-box following that line. The dark region shows the region of correlations that do not violate Uffink's inequality (but do so after the distillation protocol). As well, it follows that all correlations in this void trivialize communication complexity.}
\label{flow_map}
\end{figure}

To illustrate (see Fig. \ref{flow_map}) that we choose the two-dimension face given by the local points $L_7$ and $L_8$ and employ the wiring presented in table \ref{2dvoid_strategies} (which is the same as the example shown for the one-dimensional scenario). In this case we have $B_0^0=B_0^7=B_7^0=B_0^8=PR$, $B_7^7=B_8^8=L_7$, $B_8^7=B_7^8=L_8$, and $B_8^0=\frac{1}{2}(L_7+L_8)$.

Given a box $p=c_{0}^{(0)} PR + c_7^{(0)}L_7+c_8^{(0)}L_8$, let $c_0^{(n)}$, $c_7^{(n)}$, and $c_8^{(n)}$ be the coefficients of $PR$, $L_7$, and $L_8$ respectively, after applying $n$ wirings. We have that the output $Q_n$ of the $n$-th wiring is:
\begin{eqnarray}
\nonumber
& & q_n=\left[2-c_0^{(n-1)}-c_8^{(n-1)}\right]c_0^{(n-1)}PR+ \\  \nonumber
& & \frac{1}{2}\left[c^{(n-1)}_0c^{(n-1)}_8+ 2\left(c_7^{(n-1)}\right)^{2}+2\left(c_8^{(n-1)}\right)^{2}\right]L_7+ \\ \nonumber
& & \frac{1}{2}\left[c_0^{(n-1)}+4c_7^{(n-1)}\right]c_8^{(n-1)}L_8.
\end{eqnarray}
Notice that $c_0^{(n)}>c_0^{(n-1)}$ for all values of $c_8^{(n-1)}\neq 1$ and $c_0^{(n-1)}\neq 1$, and $\frac{c_0^{(n)}}{c_0^{(n-1)}}=2-c_0^{(n-1)}-c_8^{(n-1)}\rightarrow1$ only when $c_0^{(n-1)}\rightarrow 1$ or $c_8^{(n-1)}\rightarrow 1$, the last one cannot happen by hypothesis. Hence $c_0$ increases monotonically with $n$ implying that $q_n\rightarrow PR$ as $n\rightarrow\infty$.

\subsection{Some $3$D quantum voids are fully distillable}

Consider a generic $3$D quantum void described by the distribution
\begin{equation}
p=c_0 PR+ (1-c_0)\left( c_{i}L_{i} +c_{j}L_{j}  +c_{k}L_{k} \right) .  
\end{equation}
where we can order the coefficients as $c_i \geq c_j \geq c_k$.
Considering the $3152$ wirings such that $B^0_0=PR$ we have that there is at most one value of $i$ such that $B^0_{i}=B_0^{i}=PR$. The best one can do is to choose a wiring such that $\tilde{c}_2= c_i$. To have distillation in whole $3$D void we should have $\tilde{c}_1= c_j+c_k$ thus implying that $\tilde{c}_0= 0$. However, this is achieved only for a subset of the possible $3$D voids given by
$(L_1L_2L_3)$, $(L_1L_2L_4)$, $(L_1L_2L_5)$, $(L_1L_2L_6)$, $(L_1L_3L_4)$, $(L_1L_5L_6)$, $(L_2L_3L_4)$, $(L_2L_5L_6)$, $(L_3L_4L_7)$, $(L_3L_4L_8)$, $(L_3L_7L_8)$, $(L_4L_7L_8)$, $(L_5L_6L_7)$, $(L_5L_6L_8)$, $(L_5L_7L_8)$ and $(L_6L_7L_8)$. For all other cases it follows that $\tilde{c}_2= c_i$, $\tilde{c}_1= c_j$ thus implying that $\tilde{c}_0= c_k$, case in which the necessary condition \eqref{necessary} implies that at least the isotropic line where $c_i=c_j=c_k=1/3$ will not be distillable. 

A natural question is then whether distillable $3$D voids can all be distilled up to a PR-box. As mentioned before, a sufficient condition to achieve that is that the wiring maps the $3$D void to the same $3$D void. However, by searching over the $3152$ wirings we could not find any with this property. It could be, however, that searching over all wirings (not necessarily maping the $3$D void to another $3$D void) or considering a distillation protocols based on a higher number of copies would achieve that.

\subsection{No non-signalling face of dimension $4$ or higher is fully distillable}

The argument given above showing that not every $3$D quantum void can be fully distilled can also be extended to non-signalling faces of dimension $4$ or higher. Consider the isotropic line of a $4$-dimensional NS face
\begin{equation}
p=c_0 PR+ (1-c_0)/4\left(L_{i} +L_{j}  +L_{k}+L_{m} \right) ,  
\end{equation}
where the $4$ coefficients of the local part are the same and equal to $c_i=1/4$. The necessary criterion for distillation \eqref{necessary} implies that  $\tilde{c}_2 > \tilde{c}_0$. However, by restricting to the wirings such that $B^0_0=PR$ and making $\tilde{c}_2=c_i$, the best we can have for this scenario is $\tilde{c}_1 = c_j + c_k$, so necessarily $\tilde{c}_0=c_m$. That is, there is no wiring satisfying the necessary condition for distillation. Clearly, the argument extends to dimensions higher than $4$, thus showing that in any non-signalling face with dimension $4$ or more, at least the isotropic line of that face will not be entirely distillable.

Interestingly, the set $Q^{1}$ (the first level of the NPA hierarchy) can successfully reproduce some of the $4$D voids. At the same time, to our knowledge, there is no known  closed set of correlations between $Q^{1}$ and NS-polytope, which may be an indication that the voids reproduced by $Q^{1}$ may be asymptotically distillable to the PR-box,  if more than two copies are considered in the distillation protocol. However, for five dimensions and beyond, no asymptotic distillation to PR-box will be possible for all correlation in the quantum void. This follows from the fact that in  these cases there is always a gap between the set of quantum correlations and $Q^{1}$.

\section{nonlocality distillation and trivial communication complexity}
\label{sec:CC}
Among the several principles introduced to try to explain why correlations beyond quantum mechanics are unlikely we have the so called non-trivial communication complexity. The basic setup involves two distant parties which locally receive bit strings $\vec{x}$ and $\vec{y}$ respectively and by exchanging a limited amount of information should compute function $f(\vec{x},\vec{y})$ depending on both bit strings. It seems natural that the amount of communication required should increase with the size of the bit strings. Quantum theory is compatible with that but, as shown by Van Dam \cite{van2013implausible}, PR-boxes can make such communication complexity trivial, since their use with a single bit of exchanged information is enough to make such nonlocal computations. Later on, this result has been extended to show that any correlation achieving the value of $\mathrm{CHSH} \geq 4\sqrt{2/3} \approx 3.266$ would also trivialize communication complexity \cite{PhysRevLett.96.250401}.

Afterwards, by considering nonlocality distillation it has been shown that postquantum correlations arbitrarily close to the local set also can trivialize communication complexity \cite{brunner}. Considering the isotropic line of a $2$D quantum void given by
\begin{equation}
c_0PR+(1-c_0)/2(L_7+L_8),
\end{equation}
it has been shown in \cite{PhysRevLett.96.250401} a wiring protocol capable of distilling this distribution (after asymptotically many iterations) to a PR-box. Since a PR-box violates makes communication complexity trivial, so does this correlation. The results presented in the previous section, can thus be seen as a generalization of that. As we showed, not only the isotropic line of some $2$D quantum voids but the whole quantum void can be distilled to a PR-box, thus showing their incompatibility with the principle of non-trivial communication complexity.

\section{Tightening Information Causality with nonlocality distillation}
\label{sec:IC}
Another information-theoretical principle that has attracted considerable interest is information causality \cite{pawlowski2009information}. As in the communication complexity scenario, we have two distant parties, Alice and Bob. Alice receives a N-bit string $\vec{x} = (x_1,x_2, . . . ,x_{N})$, can send a M bits of information to Bob (where $M<N$, that is, there is bounded communication) in a message $m$ and Bob is asked to make a guess $\beta_i$ of Alice's $i$-th bit. The information causality principle basically states that  $\sum_{i=1}^{N} I (x_i : \beta_i )\leq H(M)$ ($H(M)$ being the Shannon entropy of the message and $I (x_i : \beta_i )$ the mutual information between Alice's input and Bob's guess). In other terms, the total potential information about Alice’s bit string that is accessible to Bob cannot exceed the amount of information contained in the message. Quantum correlations are in accordance with information causality but the PR-box can be shown to violate it. This can be witnessed by the violation of the Uffink's inequality, a necessary condition for a given correlation to respect information causality and given by \cite{PhysRevLett.88.230406}
\begin{equation}
(E_{00}+ E_{10})^2 +(E_{01}- E_{11})^2 \leq 4  
\label{ufink}
\end{equation} 

Interestingly, it is known that the set of correlations defined by Uffink's inequality is convex but it is not closed under wirings \cite{PhysRevA.80.062107}. That is, some correlations which do not violate Uffink's inequality can do so by nonlocality distillation \cite{PhysRevA.80.062107}. For instance, in \cite{PhysRevA.80.062107} this has been shown by considering a specific section of the non-signalling polytope and over a very small region of it. Our results can be used to extend that. As shown in \cite{PhysRevA.99.032106}, some correlations of the $2$D quantum voids do not violate Uffink's inequality (see Fig.~\ref{flow_map}). From $28$ $2$D-voids, $16$ of them has some correlations that do not violate Uffink's inequality which are: $(L_1L_3)$, $(L_1L_4)$, $(L_1L_7)$, $(L_1L_8)$, $(L_2L_3)$, $(L_2L_4)$, $(L_2L_7)$, $(L_2L_8)$, $(L_3L_5)$, $(L_3L_6)$, $(L_4L_5)$, $(L_4L_6)$, $(L_5L_7)$, $(L_5L_8)$, $(L_6L_7)$ and $(L_6L_8)$. From these $16$ sets, $4$ of them are distillable up to a PR-box, which are: $(L_1L_3)$, $(L_2L_4)$, $(L_5L_7)$ and $(L_6L_8)$. That is, by nonlocality distillation one can prove that all the correlations in these $2$D quantum voids indeed violate information causality.

\section{Discussion}
\label{sec:discussion}
In this paper we have derived a necessary condition for the nonlocality distillation of correlations in the CHSH scenario. By considering quantum voids --faces of the non-signalling set -- we have used this criterion to prove that all $1$D and some $2$D quantum voids can be distilled up to a the maximal nonlocal correlation, a PR-box. Also, we have proven that some $3$D quantum voids are fully distillable, that is, all correlations can have its nonlocality increased by wirings. However, we could not find any wirings capable of distilling correlation in $3$D void up to a PR-box, something that remains as an interesting open question. For quantum voids of dimension four or higher that is no longer possible, as the isotropic line in these voids is not distillable (at least with two copies). Interestingly, for $4D$ quantum voids and beyond, the isotropic line of these sets (defined by summing the local deterministic strategies with equal coefficients) cannot be distilled with two copies. A similar result has been proven also in the limit of infinitely many copies \cite{beigi2015monotone} for the so called isotropic-box and that can be understood as a particular case of the isotropic lines we consider here.

Building up on these results, we show the relevance of nonlocality distillation on the understanding of information-theoretical principles for quantum theory. First, we provide a generalization of the results in \cite{brunner}, showing that a large class of postquantum correlations (in $1$D and $2$D quantum voids) can make communication complexity trivial. Finally, we have shown how a distillation protocol can help understanding the set of correlations compatible with the information causality principle. More precisely, $2$D quantum void correlations that do not violate a necessary condition to respect information causality can do so after a distillation protocol. In view of all that, it would be interesting to analyze further the use of distillation protocols and in particular correlations in quantum voids to understand and characterize closed sets under wirings, in particular considering multipartite scenarios \cite{fritz2013local,PhysRevLett.107.210403,Chaves2017causalhierarchyof} to which very few principles have been introduced so far.

\acknowledgments
We acknowledge support from the Brazilian ministries MCTIC, MEC, and the CNPq (PQ Grant No. 307172$/$2017-1 and INCT-IQ and PDJ grant No. 154354$/$2018-0), the Serrapilheira Institute (Grant No. Serra-1708-15763), and from John Templeton Foundation via Grant Q-CAUSAL No. 61084.

\bibliography{Ref}

\end{document}